# Inversion and tight focusing of Airy pulses under the action of third-order dispersion.

Rodislav Driben[1,2,*], Yi Hu[3], Zhigang Chen[4], Boris A. Malomed[1], and Roberto Morandotti[3],

[1] *Department of Physical Electronics, School of Electrical Engineering, Faculty of Engineering, Tel Aviv University, Tel Aviv 69978, Israel*
2. *Department of Physics & CeOPP, University of Paderborn, Warburger Str. 100, D-33098 Paderborn, Germany*
3. *INRS-EMT, 1650 Blvd. Lionel-Boulet, Varennes, Québec J3X 1S2, Canada*
4. *Department of Physics & Astronomy, San Francisco State University, San Francisco, CA 94132*
*Corresponding author: driben@post.tau.ac.il

By means of direct simulations and theoretical analysis, we study the nonlinear propagation of truncated Airy pulses in an optical fiber exhibiting both anomalous second-order and strong positive third-order dispersions. It is found that the Airy pulse first reaches a finite-size focal area as determined by the relative strength of the two dispersion terms, and then undergoes an inversion transformation such that it continues to travel with an opposite acceleration. The system notably features tight focusing if the third-order dispersion is a dominant factor. These effects are partially reduced by a Kerr nonlinearity.

Accelerating light waves in the form of Airy beams moving along bending trajectories have recently drawn a great deal of attention [1-8]. A remarkable feature of these beams is their acceleration, an inherent property rather than a result of interactions with other waves. In addition to spatial Airy beams, the dynamics of temporal Airy pulses was also studied [9-13]. In the latter case, the acceleration of temporal pulses is characterized by their varying group velocity. Airy waves were proposed for various applications including plasmonic routing [14-15], optical light bullets [16-17], and supercontinuum generation [18]. In general, self-accelerating Airy waves are a remarkable representative of a class of waves propagating along curved trajectories [19-23].

In this Letter, we study the dynamics of truncated Airy pulses launched into a fiber close to its zero-dispersion point, under the action of the second- and third-order dispersion (TOD). In particular, we report new features driven by the dominant positive TOD, which were not reported in publication dealing with negative TOD [24].

The evolution of truncated Airy pulses, which are launched in the form of $\varphi_0 = A_0 \text{Ai}(T)\exp(aT)$ with a truncation parameter $a > 0$, obeys the normalized nonlinear Schrödinger equation:

$$i\varphi_\xi + (1/2)\text{sgn}(\beta_2)\varphi_{TT} - (1/6)i\varepsilon\varphi_{TTT} + |\varphi|^2\varphi = 0, \quad (1)$$

with $\xi = z|\beta_2|/T_0^2$ and a relative TOD strength parameter $\varepsilon = \beta_3/(|\beta_2|T_0)$, where $\beta_2$ and $\beta_3$ are the second- and third-order dispersion parameters of the fiber and $T_0$ is the pulse's width. The Kerr nonlinearity is represented by the last term. It was found that the linear propagation of the truncated Airy wave in a fiber with second-order dispersion leads to a deviation of the wave [9].

Here we report somehow surprising results when the propagation dynamics is dominated by the TOD term with $\varepsilon > 0$ in Eq. (1). The pulse reaches a focal point, then undergoes inversion, and propagates with an acceleration which is opposite in sign (Fig. 1). Past the focal point, the pulse's dynamics is similar to that of Gaussian pulses in fibers with TOD [25].

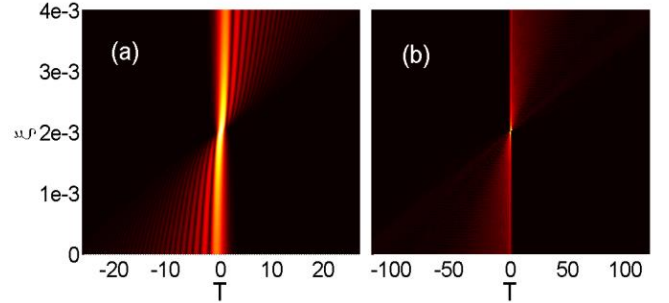

Fig. 1. (Color online) The dynamics of Airy pulses under the dominant action of TOD, with $\varepsilon = 1000$ in Eq. (1). The truncation is $a = 0.1$ in (a) and 0.01 in (b). Here and below, all results are presented in normalized units.

Hereafter we display the evolution of |$\varphi$|, rather than of the intensity |$\varphi$|$^2$, to achieve a better visualization of the dynamics of the low- intensity parts of the pulses. Figures 1(a) and 1(b) demonstrate the dynamics of the pulses in the cases of a strong and a moderate truncation, respectively. Around the focal point, all the pulse's energy is concentrated in a very short temporal slot, featuring *tight focusing* in the time domain, a property that may lead to the realization of novel pulse-compression techniques. The tight-focusing effect strongly depends on the truncation parameter, $a$. For instance, for large $a = 0.1$ the ratio between the peak powers at the focal point and in the input is about 3 [Fig. 1(a)]. Decreasing $a$, i.e., increasing the pulse energy, leads to a

much stronger energy concentration at the focus. In particular, we found that, for $a = 0.01$, the peak-power ratio is 21.48 [Fig. 1(b)], and for $a = 0.001$ it reaches the value of 37.4. Naturally, the same propagation scenario occurs under the action of a negative TOD, if the input Airy pulses are launched with a reversed acceleration.

Next, we consider the dynamics of the Airy pulses under the action of second- and third- order dispersion terms of comparable strengths. In this regime, the focal point extends to a finite area where the truncated Airy pulse experiences an acceleration reversal. The size of the area depends on the relative strength of the dispersion coefficients. Past this area, the pulse reemerges as an inverted one, and continues to propagate, see Fig.2.

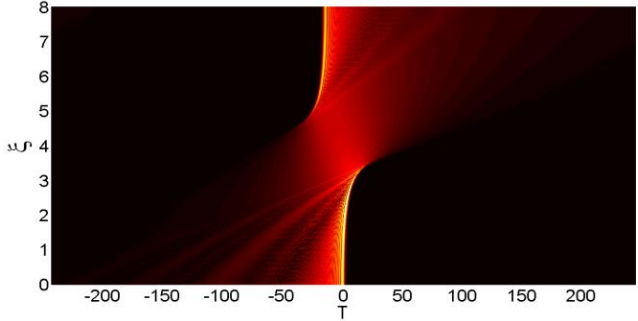

Fig. 2. (Color online) The dynamics of the truncated Airy pulse under the combined action of second- and third-order dispersions with comparable strengths, $\varepsilon = 0.5$ in Eq. (1). The truncation parameter is $a = 0.01$.

For each value of the truncation parameter, there is a minimal value of the scaled TOD coefficient, $\varepsilon_{\min}$, starting from which the inversion effect is observed. At $\varepsilon < \varepsilon_{\min}$ the pulse diverges. The dependence of $\varepsilon_{\min}$ on the initial truncation parameter, $a$, is displayed in Fig. 3(a), starting from a very small value, $a = 0.001$, up to a large one, $a = 0.1$. Figure 3(b) demonstrates the size of the reversal area, measured from one amplitude maximum to another, as a function of $\varepsilon$ for a fixed moderate truncation, $a = 0.01$. Naturally, the size diverges for very small $\varepsilon$.

The linear pulse dynamics was studied in an analytical form, using the Fourier transform of the linearized version of Eq. (1):

$$\psi_\xi = (i/2)\mathrm{sgn}(\beta_2)\omega^2\psi + (i/6)\varepsilon\omega^3\psi \quad (2)$$

In practice, an Airy pulse can be produced by adding a cubic phase to a Gaussian spectrum, i.e., solving Eq. (2) for an input $\psi_0 = \exp(-\alpha\omega^2 + i\omega^3/3)$ with $\alpha > 0$. The analytical solution is:

$$|\varphi(\xi,T)|^2 = 2\pi\theta^{2/3}\left|\mathrm{Ai}\left[-\theta^{1/3}\left(T + \frac{1}{4}\theta\xi^2 - \alpha^2\theta + i\,\mathrm{sgn}(\beta_2)\alpha\theta\xi\right)\right]\right|^2$$

$$\times \exp\left(-2\alpha\theta T + \frac{4\alpha^3}{3}\theta^2 - \theta^2\alpha\xi^2\right), \quad (3)$$

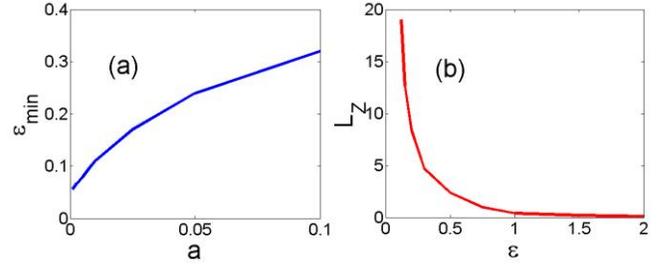

Fig. 3 (Color online) (a) The minimal scaled TOD coefficient, necessary for the occurrence of the reversal transformation, vs. the truncation parameter of the initial Airy pulse. $a$. (b) The size of the focal area vs. the TOD coefficient, at fixed $a = 0.01$.

where $\theta \equiv 1/(\varepsilon\xi/2+1)$.

This solution is valid provided that $\theta$ does not diverge, i.e., $\varepsilon\xi/2 \neq 1$. When $\varepsilon\xi/2 = 1$, the pulse profile can also be obtained from Eq. (2):

$$|\varphi(T)|^2 = \frac{1}{2\sqrt{\alpha^2 + 1/\varepsilon^2}}\exp\left(-\frac{T^2}{2\alpha + 2/(\alpha\varepsilon^2)}\right). \quad (4)$$

The intensity distribution predicted by Eqs. (3) and (4) is displayed in the inset of Fig. 4, where it is verified by comparing with direct simulations of Eq. (1) starting from the initial conditions (3). The agreement is perfect for both expressions (3) and (4).

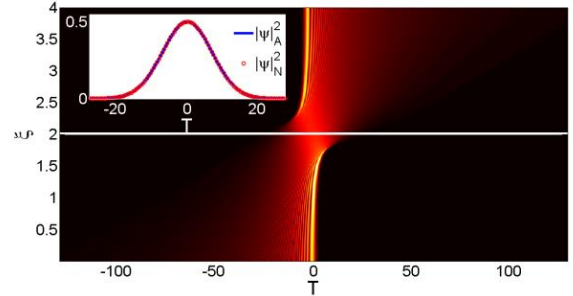

Fig. 4. (Color online) Dynamics of the truncated Airy pulse under the combined action of the second- and third-order dispersions, as predicted by Eqs. (3) and (4), for $\varepsilon = 1$ and a Gaussian-truncation parameter $\alpha = 0.02$. Inset: the intensity profile at the focal point. The solid blue line is produced by Eq. (4), while the red curve marked by circles represents direct simulations of Eq. (1).

The nonlinear term in Eq. (1) distorts the structure of the propagating Airy pulses. The analytical solutions (3) and (4) are not relevant in this case, and we solved Eq. (1) numerically by means of the well-known split-step method. The tight-focusing effect is actually reduced by the action of nonlinear self-focusing, which disrupts the phase evolution that determined the result in the linear setting, and effectively stretches the light-accumulation spot. An example of the TOD-dominated nonlinear dynamics is demonstrated in Fig. 5(a) for an initial amplitude $A_0=3$. In this case, the compression ratio is reduced to 6.77, in comparison with its counterpart 21.48 obtained above in the linear model for the same $a$, see Fig.1(b). Further, Fig. 5(b) reveals that under the action of an even stronger nonlinearity, the high-energy Airy enters a *soliton-shedding* regime, similar to that reported in Refs. [26] and [27], with the frequencies of the ejected solitons depending on the TOD strength. The subsequent trajectories of the ejected solitons are affected by multiple collisions. These collisions are inelastic due to the strong TOD, which leads to soliton frequency shifts [28].

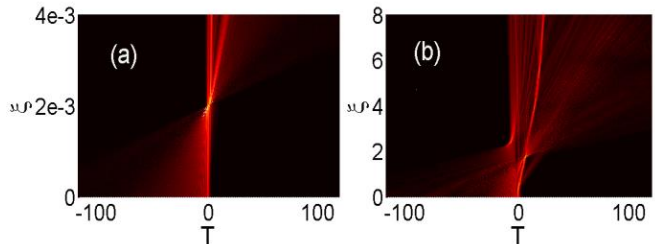

Fig. 5. (Color online) Dynamics of the Airy pulse in the nonlinear regime (a) under the prevalence of TOD [$\varepsilon =$ 1000, as in Fig. 2(b)], with $A_0 = 3$; (b) under the combined action of second and third-order dispersions with comparable strengths ($\varepsilon = 1$, as in Fig.3), with $A_0 = 5$, featuring the generation of multiple solitons. The truncation parameter is $a = 0.01$.

In conclusion, we have studied the dynamics of truncated Airy pulses in optical fibers in the presence of a strong positive TOD. When the pulse dynamics is governed by the TOD, the Airy pulse reaches the tight-focusing point, then undergoes an inversion, and finally continues to propagate with an opposite acceleration. At the focal point, the pulse is concentrated in a very narrow and intense light spot. Under the action of second- and third-order dispersion terms with comparable strengths, the focal point extends into a finite area, from which the pulse re-emerges in an inverted form. The minimal value of the TOD strength, necessary for the onset of the inversion was found as a function of the input truncation parameter. The dependence of the size of the inversion area on the TOD strength was identified as well. In the nonlinear regime, the deformation of the pulse reduces the tight-focusing effect. An even stronger nonlinearity switches the evolution of the high-energy pulse into the regime of multiple soliton generation ("shedding").

The work of R.D. and B.A.M. is supported, in part, by the Binational (US-Israel) Science Foundation through grant No. 2010239. Research is Canada is supported by the NSERQ and FQRNT funding agencies.